\documentclass[letterpaper, 10 pt, conference]{ieeeconf}  

\IEEEoverridecommandlockouts                              
\usepackage[utf8]{inputenc}
\usepackage[T1]{fontenc}

\usepackage{amsmath,amssymb,amsfonts}
\usepackage{tikz}

\title{\LARGE \bf
Investigating the Robustness of Vision Transformers against Label Noise in Medical Image Classification}

\author{Bidur Khanal$^{1}$, Prashant Shrestha$^{3*}$, Sanskar Amgain$^{3*}$, Bishesh Khanal$^{3}$, Binod Bhattarai$^{4}$ and Cristian A. Linte$^{2}$
\thanks{$^{1}$Bidur Khanal is with Center for Imaging Science, Rochester Institute of Technology, Rochester, NY, USA. \texttt{(E-mail: bk9618@rit.edu)}}
\thanks{$^{3}$Prashant Shrestha, Sanskar Amgain, and Bishesh Khanal are with NepAl Applied Mathematics and Informatics Institute for Research (NAAMII), Lalitpur, Nepal.}
\thanks{$^{4}$Binod Bhattarai is with University of Aberdeen, Aberdeen, UK.} 
\thanks {$^{2}$Cristian A. Linte is with Biomedical Engineering, Rochester Institute of Technology, Rochester, NY, USA.}
\thanks {* Equal contributions.}
}

\begin{document}

\maketitle
\thispagestyle{empty}
\pagestyle{empty}

\begin{abstract}

Label noise in medical image classification datasets significantly hampers the training of supervised deep learning methods, undermining their generalizability. The test performance of a model tends to decrease as the label noise rate increases. Over recent years, several methods have been proposed to mitigate the impact of label noise in medical image classification and enhance the robustness of the model. Predominantly, these works have employed CNN-based architectures as the backbone of their classifiers for feature extraction. However, in recent years, Vision Transformer (ViT)-based backbones have replaced CNNs, demonstrating improved performance and a greater ability to learn more generalizable features, especially when the dataset is large. Nevertheless, no prior work has rigorously investigated how transformer-based backbones handle the impact of label noise in medical image classification. In this paper, we investigate the architectural robustness of ViT against label noise and compare it to that of CNNs. We use two medical image classification datasets---COVID-DU-Ex, and NCT-CRC-HE-100K---both corrupted by injecting label noise at various rates. Additionally, we show that pretraining is crucial for ensuring ViT's improved robustness against label noise in supervised training.

\end{abstract}

\section{INTRODUCTION}

Label noise in medical classification datasets can arise from several factors, including inter-observer variability during annotation \cite{radsch2023labelling}, the use of non-expert annotators to reduce costs \cite{orting2020survey}, and the growing reliance on automated labeling algorithms \cite{irvin2019chexpert}. It is well established that deep learning-based supervised medical image classification requires accurately annotated labels to effectively train classification models. Supervised training with inaccurately annotated or noisy labels can impair a model's generalizability, resulting in subpar test performance \cite{lee2019robust,zhang2021understanding,khanal2021does, khanal2023investigating}.

In response, recent studies have focused on training models to be robust against label noise in medical image classification, encompassing a wide range of datasets, including skin cancers, breast tumors, thoracic diseases, chest infections, retinal diseases, and prostate cancers. These approaches incorporate various techniques, such as label smoothing \cite{pham2021interpreting}, estimating a label noise transition matrix to modify end layers \cite{dgani2018training}, sample re-weighting \cite{le2019pancreatic,xue2019robust}, consistency regularization \cite{zhou2023combating}, employing student-teacher networks \cite{xue2022robust}, or relying on self-supervised pretraining \cite{khanal2023improving}. Each technique offers unique advantages, depending on the nature of the noise and the characteristics of the dataset. A common element in all these methods is the use of CNN-based backbone networks, a trend also evident in state-of-the-art general Learning with Noisy Label (LNL) methods \cite{han2018co, Li2020DivideMix:,li2022selective}.

Vision Transformers (ViTs) \cite{dosovitskiy2020image} have recently gained popularity, outperforming CNNs in numerous benchmarks across both computer vision and medical datasets \cite{han2022survey,shamshad2023transformers}. A key feature of transformers is their inherent attention mechanism, which adeptly captures long-range dependencies across different spatial regions of an image, thereby providing a comprehensive global context \cite{dosovitskiy2020image}. Transformers provide greater flexibility in learning, in contrast to CNNs, which primarily focus on local context. Despite these advantages, to our knowledge, there has been no recent work on LNL using ViT as the backbone for medical classification tasks containing noisy training labels. This raises an important question: How effective are ViTs in handling label noise in medical image classification, and what is their robustness in such scenarios?

In this study, we examine the resilience of ViT against label noise in medical image classification. We use two publicly available datasets for our investigation: i) COVID-DU-Ex \cite{tahir2021covid}, a chest X-ray infection classification dataset, and ii) NCT-CRC-HE-100K \cite{kather2019predicting}, a histopathology tissue classification dataset. We also explore the application of two self-supervised pretraining techniques on ViTs for these datasets to enhance robustness against label noise. Furthermore, to assess the benefit of a ViT vs. a CNN backbone, we compare the performance of a well-known LNL method—Co-teaching \cite{han2018co}, which typically trains robustly even with label noise, by replacing its CNN backbone with a ViT. This allows us to evaluate how the performance differs from the conventional CNN-based system. Our study does not introduce new methods; rather, it aims to shed light on the architectural robustness of ViTs relative to CNNs, against label noise in medical image classification, a topic that needs more in-depth investigation.

\section{METHODOLOGY}

\subsection{Label Noise}
\label{label_noise}
Let us consider a dataset $D = \{(\mathbf{x}_i, y_i)\}_{i}^{n}$ comprising $n$ samples, where $x_i \in \mathbb{R}^{d}$ is an input and $y_i \in \{1,2,3,...,c\}$ is its corresponding true label. To mimic a noisy real-world dataset, we synthetically inject label noise into this clean dataset by randomly flipping the true label $y_i$, to an incorrect label $\hat{y_{i}}$ with a certain probability $p$, also referred to as label noise rate. Here, $\hat{y_{i}}$ can be any class label from the dataset, except for the true label, i.e., $\hat{y_{i}} \overset{p}{\sim} \{1,2,3,...,c\}\setminus\{y_{i}\}$.

\subsection{Dataset}

We experimented with two datasets. The COVID-DU-Ex dataset \cite{tahir2021covid} consists of $33,920$ chest X-ray images, classified into COVID, non-COVID, and normal categories. Out of these, $27,132$ images were used for training, and the remaining $6,788$ images comprised the test set. The NCT-CRC-HE-100K dataset \cite{kather2019predicting} includes histopathology image patches of nine different tissue types, with a total of $100,000$ images for training. It also has a separate test set comprising $7,180$ samples from the same nine categories. We maintained the size of inputs to $224 \times 224$ for both datasets across all the experiments.

\subsection{Training with Noisy Labels}

\begin{figure}[t!]
\centering
\includegraphics[width=1\linewidth]{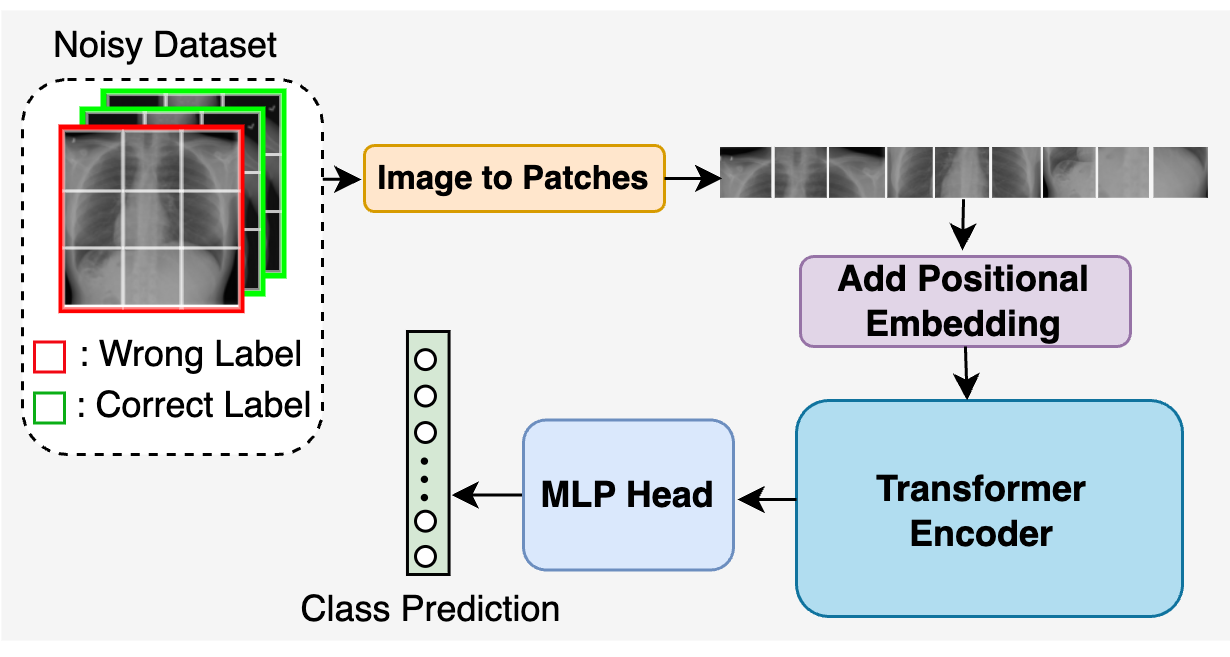}
\vspace{-0.5em}
    \caption{A pipeline for training medical image classification task with label noise, where the feature extractor backbone is a transformer.}
     \label{fig:pipeline}
\end{figure}

As described in Section \ref{label_noise}, we injected label noise at various rates \( p = \{0, 0.1, 0.2, 0.3, ..., 0.9\} \) into both datasets. Subsequently, we trained ViT for image classification task using the noisy datasets (see Fig. \ref{fig:pipeline}). Unlike CNNs, ViTs divides the input image into fixed-sized patches, which are then flattened into vectors. The flattened vectors are concatenated with positional embeddings and fed into a transformer encoder, followed by an MLP for class predictions.

We initially trained ViTs and a CNN-based model (ResNet18) using standard cross-entropy loss, which lacks inherent robustness against label noise. Then, we employed Co-teaching, a LNL method, designed to mitigate the impact of label noise. This method selects clean samples based on training loss, thereby avoiding training on inaccurately labeled samples. While Co-teaching traditionally uses CNNs as the backbone, we also experimented by replacing the backbone with ViT. To keep our investigation unbiased and focus solely on understanding the inherent robustness of transformers against label noise, we first trained all methods from scratch, avoiding the influence of models pretrained on large external natural-image datasets.

However, in the next phase, we pretrained the models using self-supervised techniques on the respective datasets to investigate whether pretraining improves the robustness of ViT against label noise. It is well known that self-supervised pretraining enhances robustness against label noise in CNN-based models \cite{khanal2023improving}. Since self-supervised learning does not rely on given labels for training, it is not impacted by label noise, thereby learning more robust feature representations. To test this hypothesis with a ViT, we utilized two self-supervised techniques, MAE~\cite{he2022masked} and SimMIM~\cite{xie2022simmim}, to pretrain on their respective datasets.

Both MAE and SimMIM rely on the reconstruction of masked image patches to learn the representations in an unsupervised manner. While MAE utilizes an encoder-decoder framework with Mean Squared Error (MSE) to regress the pixel predictions, SimMIM is an encoder only framework with a simple prediction head and utilizes $L_1$ distance to supervise the reconstruction. Consequently, we repeated all cross-entropy and Co-teaching experiments using ViTs pretrained with MAE and SimMIM.

\subsection{Evaluation}
We evaluate classification performance using balanced accuracy, measuring both the best test accuracy across all epochs (BEST) and the average test accuracy of the last five epochs (LAST). The best test accuracy indicates the maximum performance achieved by the model, while the average of the last five epochs reflects whether the model overfits the noisy labels in the training data \cite{khanal2023improving}.

\section{Implementaion Details}


We experimented with two ViT configurations: ViT Base and ViT Small. ViT Base is larger with more parameters, while ViT Small is more compact. Both models use $12$ layers, have an MLP ratio of $4$, and employ $16\times16$ patches. The ViT Small model has $6$ heads and an encoder dimension of $384$, while the larger ViT Base model has $12$ heads with an encoder dimension of $768$. For CNN, we selected the ResNet18 architecture. In the following subsections, we discuss details for supervised training with noisy datasets, followed by self-supervised pretraining for the ViTs.

\subsection{Supervised Training with Noisy labels}
We use two approaches for supervised training with noisy labels: the first uses standard cross-entropy loss without any modifications, while the second uses Co-teaching, an LNL method, for robust training with label noise.

\subsubsection{Standard cross-entropy}
We adopted the same training settings for both the ViT Small and ViT Base models across both datasets. The data augmentation was limited to basic techniques, including random horizontal flips and rotations up to $10^\circ$. We utilized the AdamW optimizer ($\beta_1=0.9,\beta_2=0.95$), with weight decay of $1e^{-4}$, an initial learning rate of $3e^{-5}$, and a Cosine Annealing learning rate scheduler. The models were trained for 50 epochs using a batch size of 128, a duration sufficient for the learning curve to saturate. For the ResNet18 model, we followed a similar 50-epoch training regimen but with a batch size of 256. We employed an SGD optimizer with a momentum of 0.9 and a weight decay of $1e^{-4}$. The initial learning rate was set to 0.01, while using Cosine Annealing scheduler.

\subsubsection{Co-teaching}
We applied the same general training settings as used in standard cross-entropy for both the ViT models and ResNet18 across both datasets. For Co-teaching, the warm-up epochs were set to $5$ for ViT and $10$ for ResNet. The method-specific hyperparameters were set as $\tau = p$ and $c = 1$, where $p$ represents the label noise rate. In our experiments, we tested label noise rates of $p = \{0.4,0.5,0.6,0.7\}$ for the COVID-DU-Ex dataset and $p = \{0.4,0.5,0.6,0.7, 0.8\}$ for the NCT-CRC-HE-100K dataset, representing the typical range of high label noise. Co-teaching cannot enhance performance beyond this range, as the noise rate surpasses a critical tipping threshold ($c-1/c$, where $c$ is the number of classes) \cite{oyen2022robustness}.

All the experiments were trained using PyTorch 12.1 in Python 3.8, using A100 GPUs. Each experiment involving training with noisy labels was repeated three times with random seeds, and the results were averaged for evaluation.

\subsection{Pretraining}
For MAE, we set the mask ratio to $75\%$ of the image patches. The image decoder utilizes a 6-layer transformer with an embedding dimension set to $512$. For data augmentation, we employed randomly resized crops with a scale ranging from $0.2$ to $1.0$. We utilized the AdamW optimizer ~\cite{loshchilov2017decoupled} ($\beta_1=0.9, \beta_2=0.95$) along with a Cosine learning rate scheduler, setting the learning rate to $1.5e-4$ and weight decay to $0.05$. The models were trained until convergence using a batch size of $128$. For COVID-DU-Ex, we trained for up to $800$ epochs for both ViT Base and ViT Small, while for NCT-CRC-HE-100K, the training was done for up to $600$ epochs. The same training configurations were used for both datasets, except for the number of training epochs.

For SimMIM, $60\%$ of the image patches were masked. We used the AdamW optimizer ($\beta_1=0.9, \beta_2=0.95$) along with a Cosine learning rate scheduler, a base learning rate of $1e-4$, a weight decay of $0.05$, and warm-up epochs of $10$. The models were trained using batch sizes of $512$ for ViT Small and $256$ for ViT Base, respectively, until the loss converged. For COVID-DU-Ex, training was done for up to $800$ epochs for both ViT Base and ViT Small. For NCT-CRC-HE-100K, ViT Small was trained for up to $800$ epochs, and ViT Base for up to $400$ epochs. We used the same training configurations for both datasets, except for the number of training epochs.

\section{RESULTS}
\subsection{Quantitative Results}
In this section, we quantitatively evaluate supervised training in both datasets at various noise rates. We first compare ViTs to ResNet18 without pretraining, then assess the impact of self-supervised pretraining, and finally compare Co-teaching with ViTs as the backbone to ResNet18.

\subsubsection{ Architecture's Role in Noisy Label Training} 
\begin{figure}[h!]
\centering
\includegraphics[width=1\linewidth]{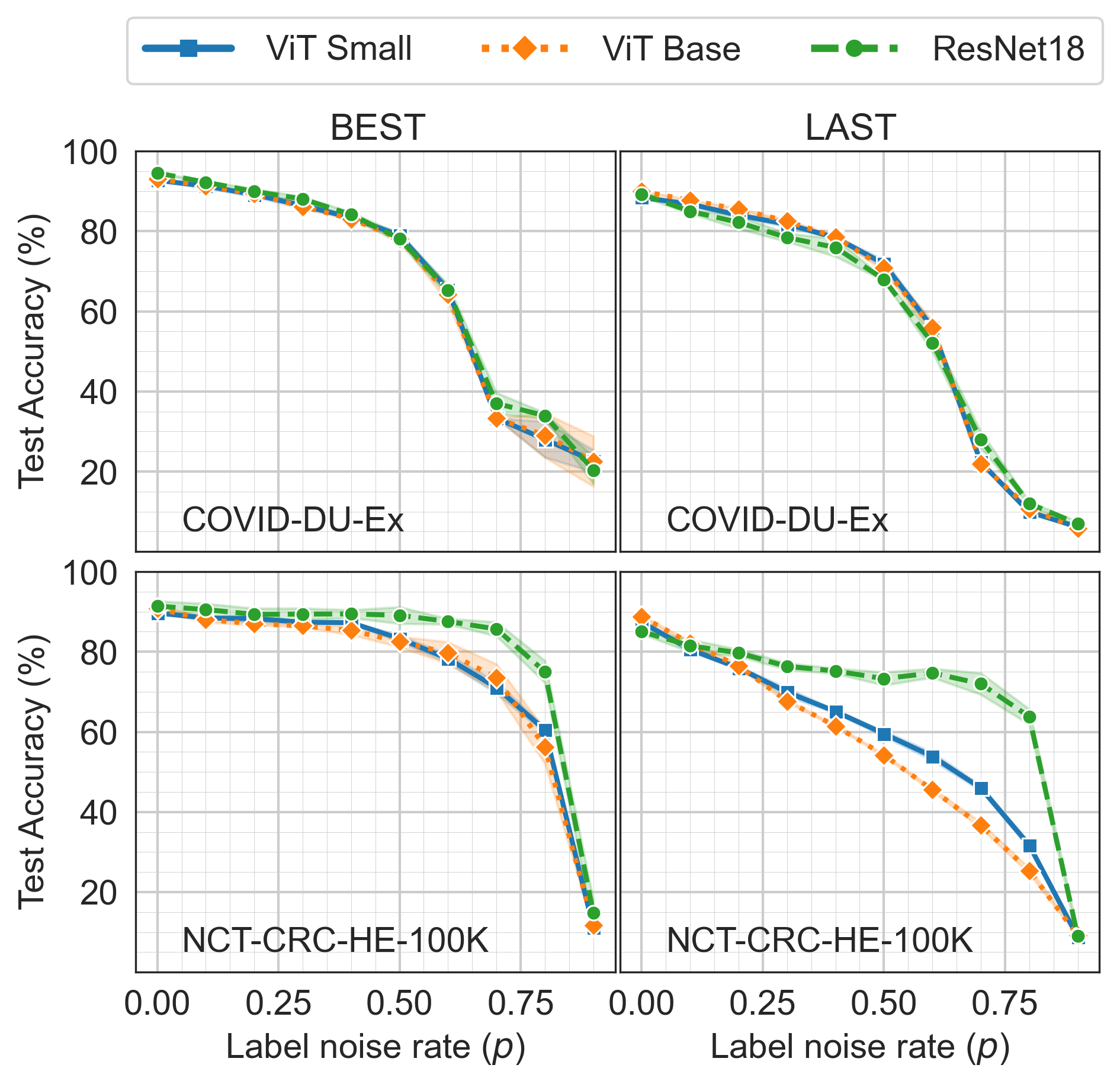}
\caption{Performance comparison of ViT Small, Vit Base and ResNet18 trained with label noise rates (p) across COVID-DU-Ex and NCT-CRC-HE-100K dataset. BEST represents the peak test performance across all epochs and LAST represents the average of test performance in the last five epochs.} 
\label{fig:standard_cross_entropy_scratch}
\end{figure}

To assess the tolerance of the architecture against label noise, we compared ViTs and ResNet18 trained from scratch with standard cross-entropy loss in Fig. \ref{fig:standard_cross_entropy_scratch}. Both ViTs and ResNet18 exhibit declining test accuracy as label noise rates increase in both the COVID-DU-Ex and NCT-CRC-HE-100K datasets, with a similar trend observed in both. While ViTs and ResNet18 perform similarly in the COVID-DU-Ex dataset, ResNet18 outperforms ViTs in the NCT-CRC-HE-100K dataset. We also conducted a t-test using the BEST score to determine whether ViT Small and ResNet18 performances are statistically different. In the NCT-CRC-HE-100K, the difference between ViT Small and ResNet18 is statistically significant with a $p$-value < $0.05$ for $p = \{0.5, 0.6, 0.7, 0.8\}$.\footnote{The $p$-value denotes statistical significance, whereas $p$ represents label noise rates.} In contrast, for COVID-DU-Ex, the difference is statistically insignificant with a $p$-value > $0.05$, further validating the results seen in the graph. Notably, ViT Base LAST's performance for NCT-CRC-HE-100K is worse than the corresponding BEST performance, potentially due to ViT Base's large parameter size and greater flexibility, making it more prone to overfitting noisy labels.

\begin{figure}[h!]
\centering
\includegraphics[width=1\linewidth]{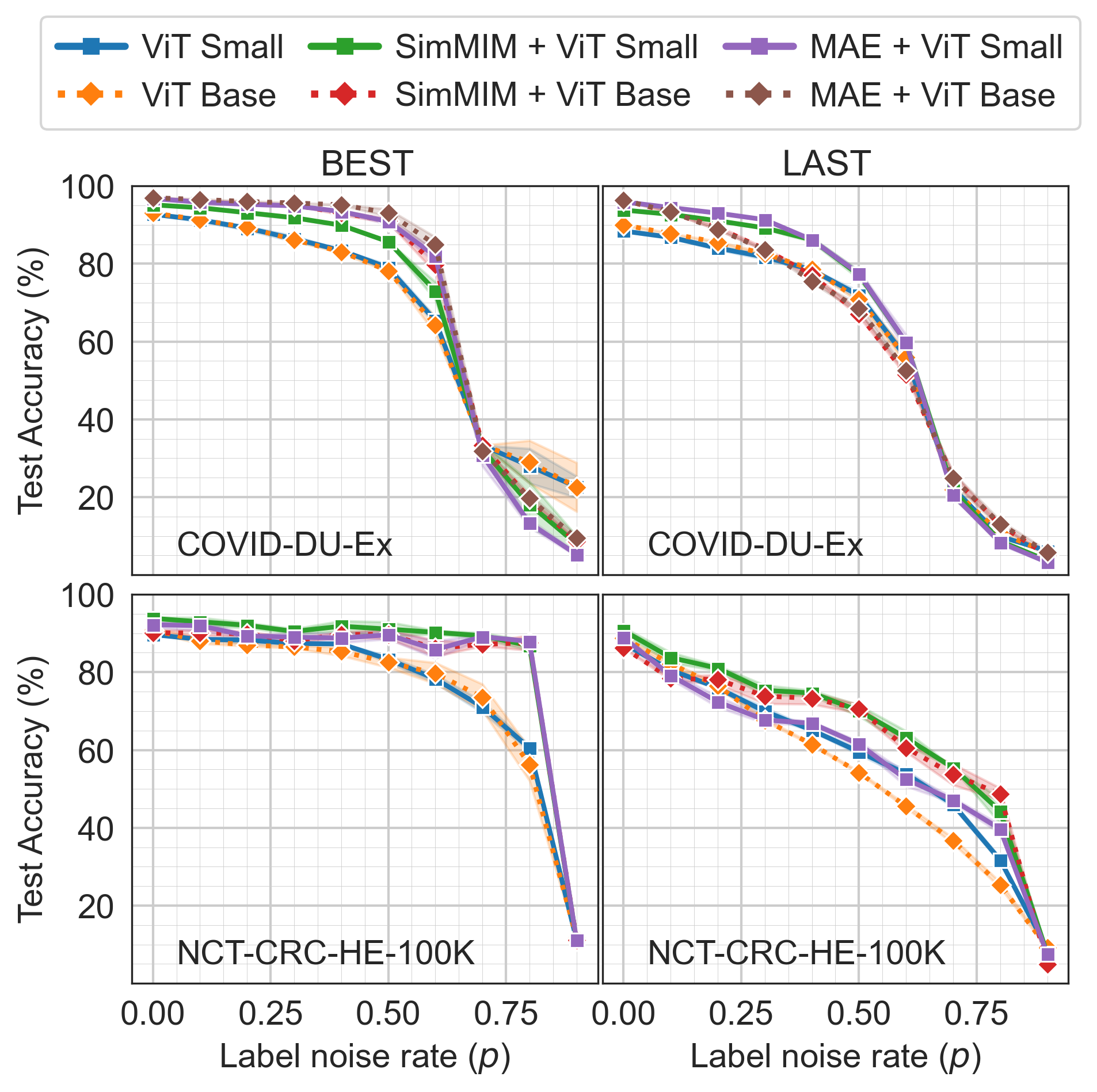}
\caption{Performance comparison of ViT Small, Vit Base at various noise rates in COVID-DU-Ex and NCT-CRC-HE-100K dataset, without/with pretraining using MAE and SimMIM.BEST represents the peak test performance across all epochs and LAST represents the average of test performance in the last five epochs.} 
\label{fig:standard_cross_entropy_pretrained}
\end{figure}

\subsubsection{Influence of Self-supervised Pretraining in ViTs}
Supervised training with ViTs benefits from self-supervised pretraining -- which improves the learned representation. In Fig. \ref{fig:standard_cross_entropy_pretrained}, we compare the impact of two self-supervised pretraining methods, MAE and SimMIM, on label-noise tolerance during supervised training with standard cross-entropy. The results indicate that pretraining improves performance for both datasets, especially in the BEST performance metric, and this improvement is consistent for both MAE and SimMIM. The performance boost is more pronounced at high label noise levels in NCT-CRC-HE-100K compared to COVID-DU-Ex. However, we observe that even after pretraining, ViT Base begins to overfit to noisy labels during training, as indicated by the lower LAST performance, potentially due to it being a large model and more susceptible to overfitting on label noise.

\begin{figure}[h!]
\centering
\includegraphics[width=1\linewidth]{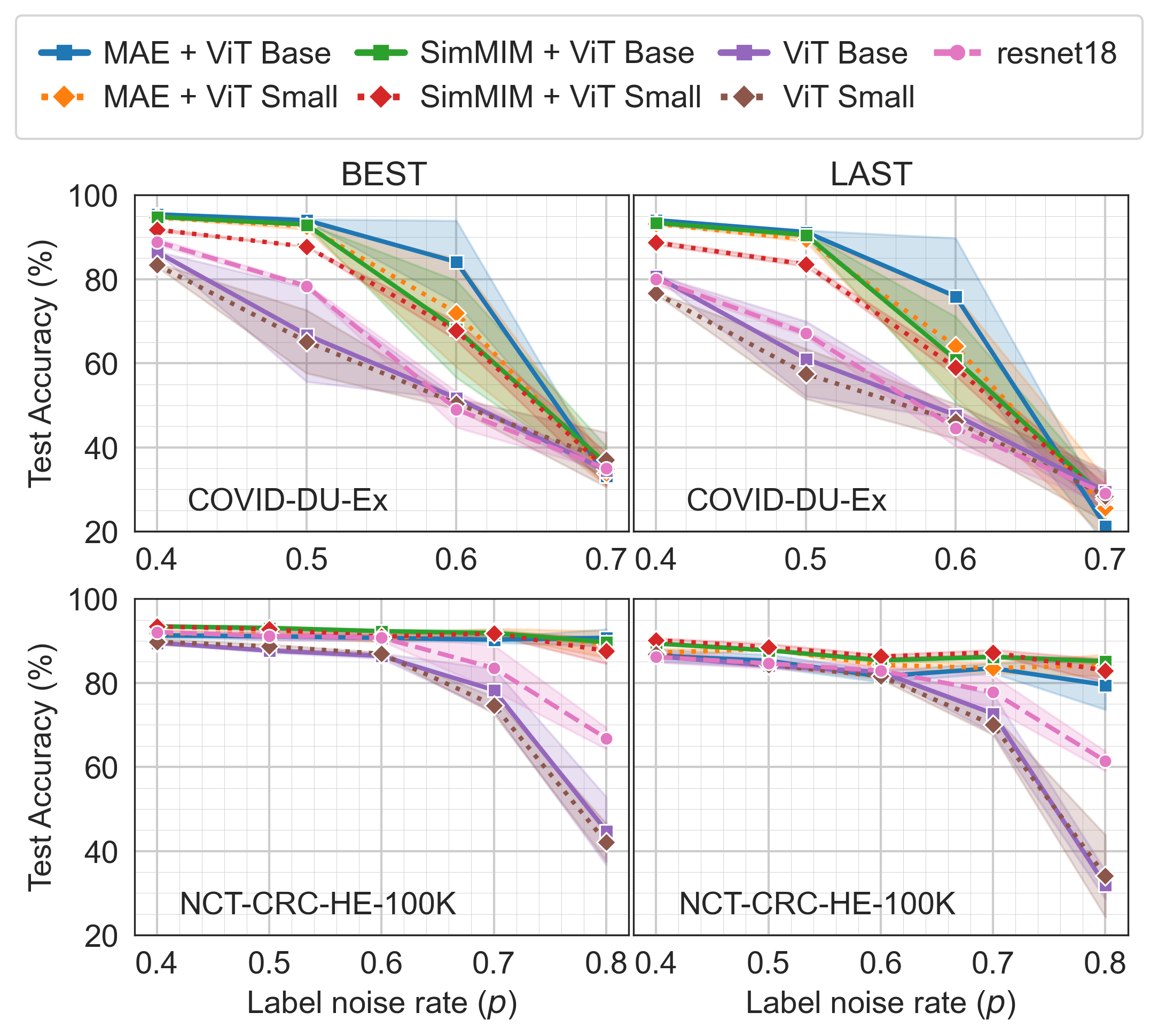}
\caption{Comparison of Co-teaching performance across various architectures at different noise rates in COVID-DU-Ex and NCT-CRC-HE-100K dataset, without/with pretraining using MAE and SimMIM. BEST represents the peak test performance across all epochs and LAST represents the average of test performance in the last five epochs.} 
\label{fig:coteaching_curve}
\end{figure}

\subsubsection{Co-teaching with ViT}

We also compared how ViTs perform when used with the LNL method, such as Co-teaching (see Fig. \ref{fig:coteaching_curve}). Co-teaching does not perform well with ViTs when trained from scratch. In comparison, the performance of Co-teaching with ResNet-18 is not relatively worse than that with ViT Base and ViT Small trained from scratch. However, when transformers are pretrained, both ViT Base and ViT Small show significant performance improvements, as confirmed by the t-test results. Specifically, MAE+ViT Small significantly surpasses ViT Small in performance, with a $p$-value < 0.05 across all noise levels ($p$) for NCT-CRC-HE-100K, and at $p = \{0.4, 0.5, 0.6\}$ for COVID-DU-Ex. These results suggest that when using ViT as a backbone instead of CNN, pretraining is crucial for achieving label-noise tolerance during training.

\subsection{Qualitative Results}

In this section, we present qualitative results by examining attention maps and prediction outcomes for selected test samples from both COVID-DU-Ex and NCT-CRC-HE-100K.

\begin{figure}[h!]
\centering
\includegraphics[width=1\linewidth]{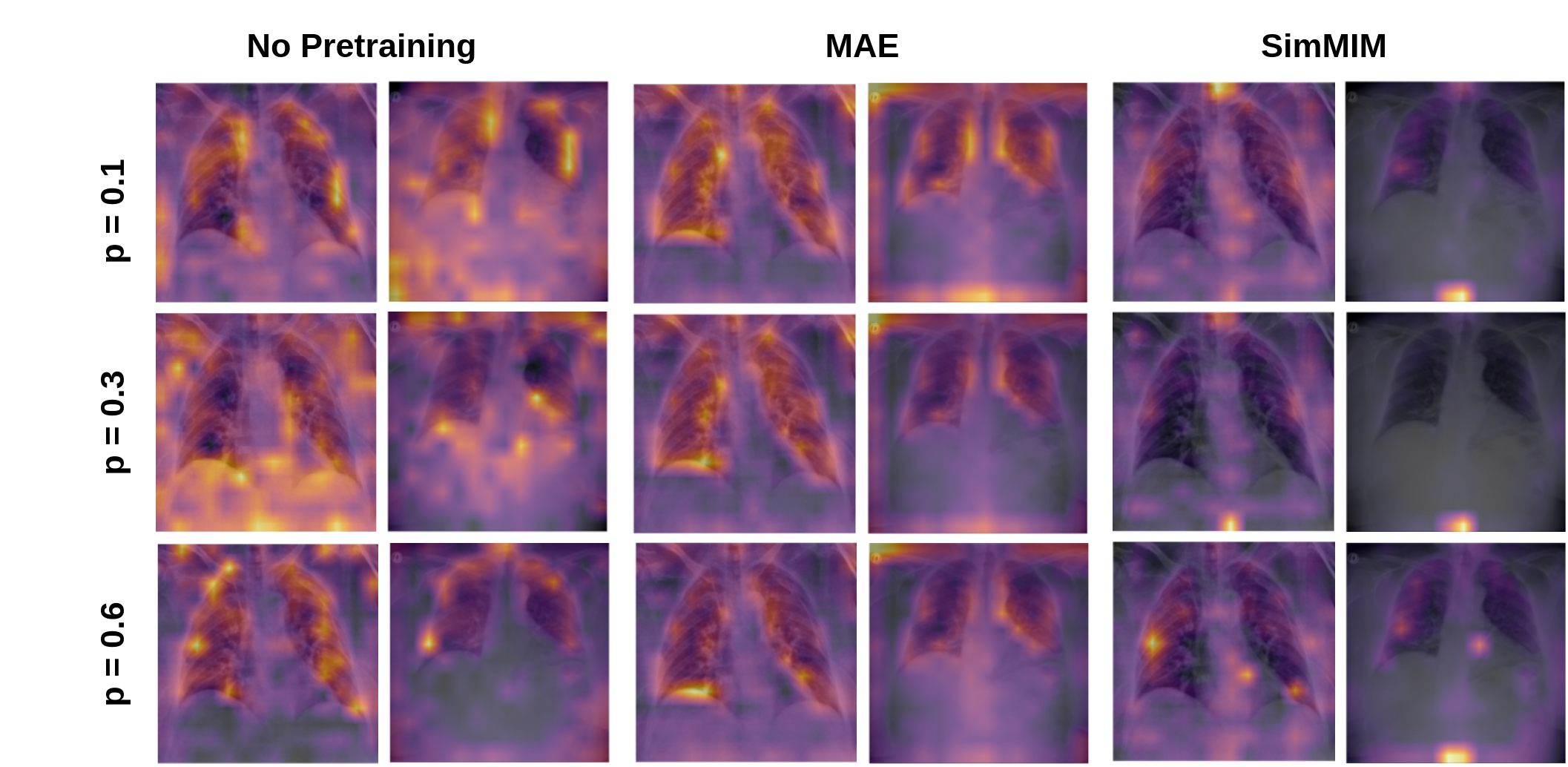}
\caption{Attention maps of ViT Small for COVID class in COVID-DU-Ex dataset, featuring two randomly selected test samples. Rows represent models trained with different label noise rates $p = (0.1, 0.3, 0.6)$. Columns 1-2: No pretraining, Columns 3-4: MAE pretraining, Columns 5-6: SimMIM pretraining before supervised training with standard cross-entropy loss on noisy labels. } 
\label{fig:attention_map_covid}
\vspace{-1em}
\end{figure}

\begin{figure}[h!]
\centering
\includegraphics[width=1\linewidth]{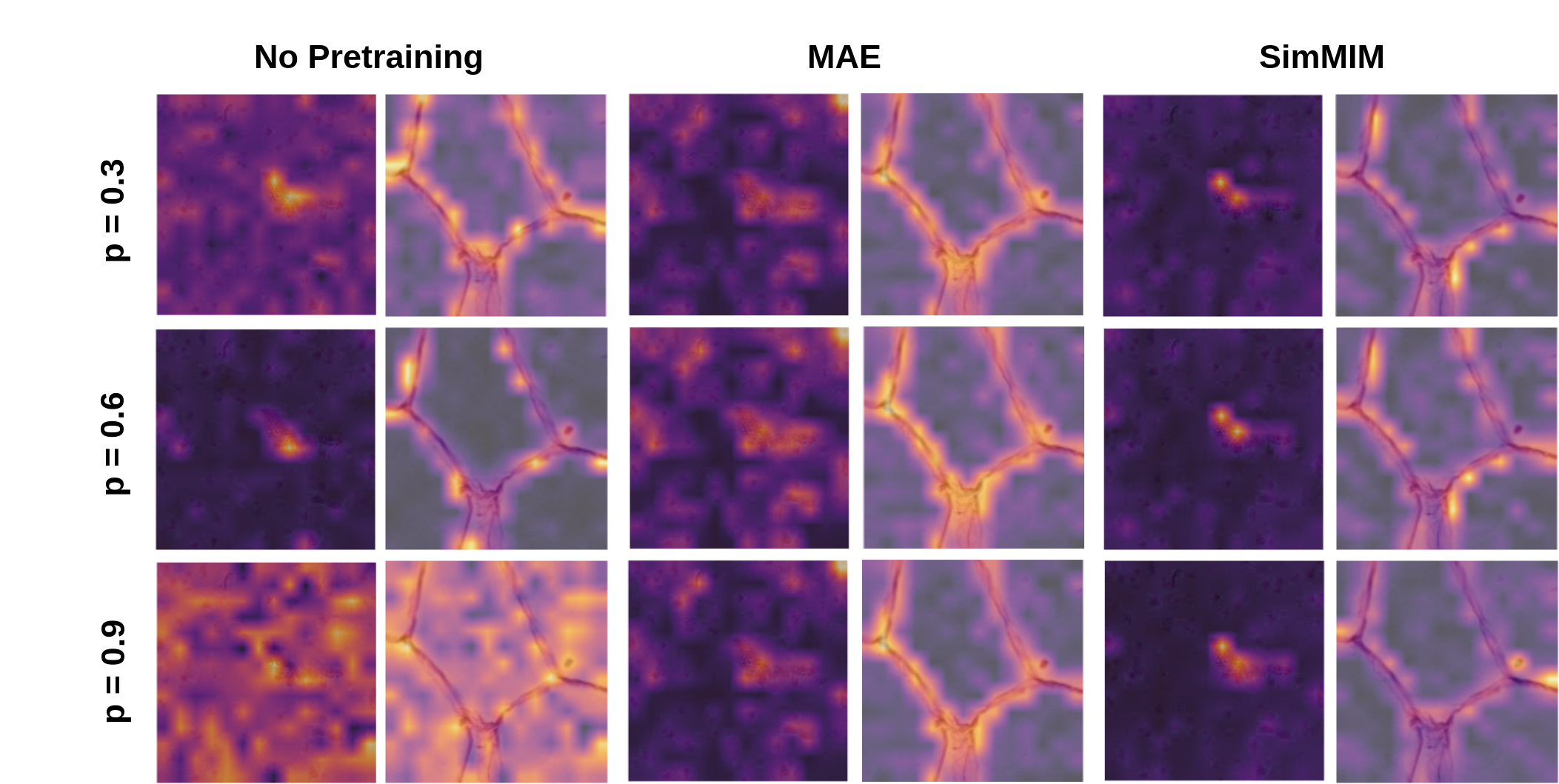}
\caption{Attention maps of ViT Small for Background (left) and Adipose (right) classes in NCT-CRC-HE-100K dataset, featuring two randomly selected test samples. Rows represent models trained with different label noise rates $p = (0.1, 0.3, 0.6)$. Columns 1-2: No pretraining, Columns 3-4: MAE pretraining, Columns 5-6: SimMIM pretraining before supervised training with standard cross-entropy loss on noisy labels. } 
\label{fig:attention_map_histopathology}
\end{figure}

\begin{figure*}[ht!]
\centering
\includegraphics[width=1\linewidth]{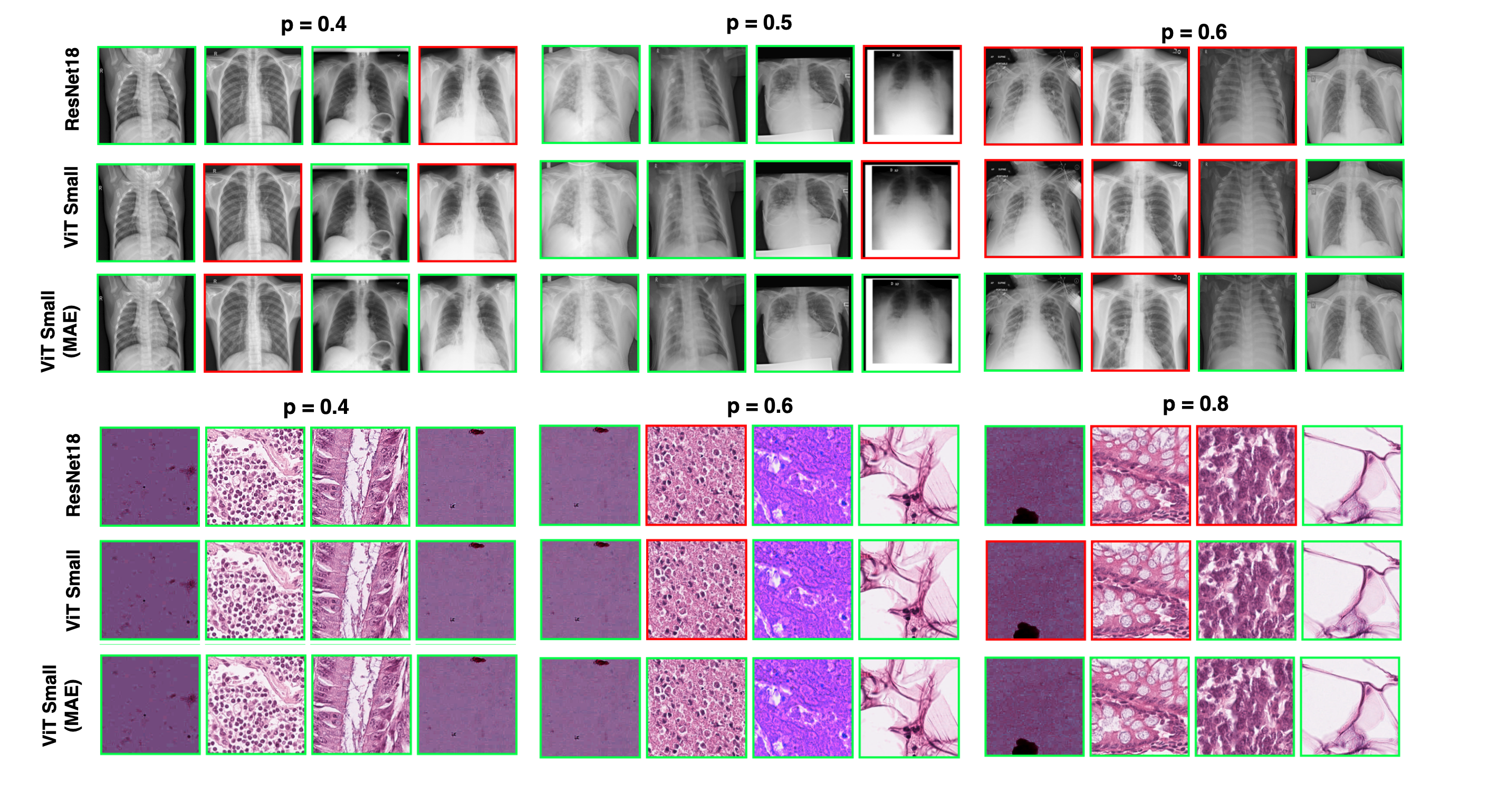}
\caption{Qualitative Results: Predictions on representative test samples from COVID-DU-Ex and NCT-CRC-HE-100K datasets using ResNet18 trained with Co-teaching, ViT Small trained with Co-teaching (not pretrained and pretrained with MAE) with noisy labels at three noise rates ($p$). Red boxes indicate incorrect predictions, and green boxes indicate correct predictions.} 
\label{fig:coteaching_prediction}
\end{figure*}

First, we analyze the attention maps generated by the trained ViT Small to investigate the potential impact of training with label noise. These attention maps are computed using Attention Rollout \cite{abnar2020quantifying}, with attention weights averaged across all heads and attention matrices multiplied between layers to produce a single map per image.

Fig. \ref{fig:attention_map_covid} and Fig. \ref{fig:attention_map_histopathology} depict the attention map visualizations for selected test samples from the COVID-DU-Ex and NCT-CRC-HE-100K datasets, respectively. Notably, ViT Small, when trained on noisy labels without pretraining, tends to produce attention maps with greater noise and dispersion, particularly evident in the COVID-DU-Ex samples. In contrast, ViT Small pretrained with MAE tends to generate cleaner maps focused on the lung regions (Fig. \ref{fig:attention_map_covid}). In the context of the NCT-CRC-HE-100K dataset, employing both MAE and SimMIM for pretraining enhances the quality of the attention maps in ViT Small trained with noisy labels, as shown in Fig. \ref{fig:attention_map_histopathology}. For example, the attention map for the \textit{Adipose} class reveals that the model is correctly focusing on the cytoplasmic membrane to distinguish the class. Interestingly, in the NCT-CRC-HE-100K samples, attention maps from the ViT Small without pretraining become increasingly noisier as training label noise increases, while those from the pretrained ViT remain relatively unchanged. However, we do not observe a similar deterioration in the attention maps from the ViT without pretraining in COVID-DU-Ex, as the training noise increases. 


In Fig. \ref{fig:coteaching_prediction}, we analyze the prediction outcome by the models trained with Co-teaching on randomly selected representative samples from both the COVID-DU-Ex and NCT-CRC-HE-100K datasets. We compare the test performance of three models: ResNet18, ViT Small (without pretraining), and ViT Small (pretrained with MAE), trained at various label noise rates. The pretrained ViT Small predicts samples correctly even in the presence of high label noise, outperforming ResNet18 and ViT Small without pretraining. However, at a low noise rate of $0.4$ for the NCT-CRC-HE-100K dataset, all the backbones appear to consistently yield accurate predictions, whether pretrained or not.

%


\section{CONCLUSION}

In this study, we explored the relative robustness of ViTs against label noise in medical image classification compared to CNN-based architectures like ResNet18. Our results suggest that ViTs are more susceptible to overfitting, particularly with larger model sizes. Without pretraining, ViTs are less effective than basic CNN-based architectures for LNL methods. However, if ViTs undergo proper pretraining using self-supervised methods before being applied with LNL methods, their robustness against label noise significantly improves. Therefore, proper pretraining is crucial for employing ViT as the backbone in LNL methods to enhance robustness against label noise.

\section*{ACKNOWLEDGMENT}
This research was supported by the NIGMS Award No. R35GM128877 of the National Institutes of Health, and OAC Award No. 1808530 and CBET Award No. 2245152, both of the National Science Foundation. We also acknowledge Research Computing at the Rochester Institute of Technology \cite{RITRC} for providing computing resources.


\bibliographystyle{plain}
\bibliography{mybib}

\end{document}